\documentclass[aps,prb,amsmath,amssymb,twocolumn,floatfix,showpacs,superscriptaddress]{revtex4-1}         
\pdfoutput=1
\usepackage[usenames,dvipsnames,svgnames,table]{xcolor}
\usepackage{amsmath,amssymb,mathrsfs,bm,feynmf,setspace}
\usepackage{graphicx}     
\usepackage[tight]{subfigure}       
\usepackage{color} 
\usepackage{lipsum}   
\usepackage{natbib}
\usepackage{braket}  
\usepackage[colorlinks=true]{hyperref}         
\hypersetup{
    bookmarks=true,         
    unicode=false,          
    pdftoolbar=true,        
    pdfmenubar=true,        
    pdffitwindow=false,     
    pdfstartview={FitH},    
    pdftitle={Cubic trihedral corner entanglement for a free scalar},    
    pdfauthor={Author},     
    pdfsubject={Subject},   
    pdfcreator={Creator},   
    pdfproducer={Producer}, 
    pdfkeywords={keyword1} {key2} {key3}, 
    pdfnewwindow=true,      
    colorlinks=true,       
    linkcolor=magenta, 
    citecolor=blue,        
    filecolor=magenta,      
    urlcolor=cyan           
} 
\definecolor{darkblue}{rgb}{0.2, 0, 0.8}
\definecolor{darkgreen}{rgb}{0.2, 0.71, 0}   
\makeatletter
\newcommand\reffnmark{%
    \@ifstar{\@reffnmarks}{\@reffnmark}}
\newcommand{\@reffnmarks}[1]{%
    \begingroup
        \unrestored@protected@xdef\@thefnmark{\ref{#1}}%
    \endgroup
    \@footnotemark}
\newcommand{\@reffnmark}[1]{(\ref{#1})}
\makeatother

\numberwithin{equation}{section}

\DeclareMathOperator{\Tr}{Tr}
\newcommand{\Renyi}{R\'{e}nyi\ }
\newcommand{\req}[1]{(\ref{#1})} 
\newcommand{\labell}[1]{\label{#1}}

\newcommand{\bea}{\begin{eqnarray}}
\newcommand{\eea}{\end{eqnarray}}
\newcommand{\ba}{\begin{eqnarray}}
\newcommand{\ea}{\end{eqnarray}}

\newcommand{\beq}{\begin{equation}}
\newcommand{\eeq}{\end{equation} }  
\newcommand{\beqa}{\begin{eqnarray}}
\newcommand{\eeqa}{\end{eqnarray}}
\newcommand{\beqar}{\begin{eqnarray*}}
\newcommand{\be}{\begin{equation}}
\newcommand{\ee}{\end{equation}}
\newcommand{\eeqar}{\end{eqnarray*}}

\newcommand{\reef}[1]{(\ref{#1})}
\newcommand{\ssc}{\scriptscriptstyle}
\newcommand{\eg}{{\it e.g.,}\ }
\newcommand{\ie}{{\it i.e.,}\ }

\newcommand{\mt}[1]{\textrm{\tiny #1}}

\newcommand{\veps}{\varepsilon}

\newcommand{\cO}{\mathcal{O}}

\newcommand{\cA}{{\cal A}}

\newcommand{\ren}{R\'enyi\ }

\newcommand{\see}{S_{\rm \ssc EE}} 

\newcommand{\al}{\alpha}

\renewcommand{\href}[2]{#2}

\begin{document}

\title{Cubic trihedral corner entanglement for a free scalar}     

\author{Lauren E. Hayward Sierens}
\affiliation{Department of Physics and Astronomy, University of Waterloo, Ontario, N2L 3G1, Canada}
\affiliation{Perimeter Institute for Theoretical Physics, Waterloo, Ontario N2L 2Y5, Canada}

\author{Pablo Bueno}
\affiliation{ Instituut voor Theoretische Fysica, KU Leuven, Celestijnenlaan 200D, B-3001 Leuven, Belgium}

\author{Rajiv R. P. Singh}
\affiliation{Department of Physics, University of California Davis, CA 95616, USA}

\author{Robert C. Myers}
\affiliation{Perimeter Institute for Theoretical Physics, Waterloo, Ontario N2L 2Y5, Canada}

\author{Roger G. Melko}
\affiliation{Department of Physics and Astronomy, University of Waterloo, Ontario, N2L 3G1, Canada}
\affiliation{Perimeter Institute for Theoretical Physics, Waterloo, Ontario N2L 2Y5, Canada}

\date{\today}
\begin{abstract}     
We calculate the universal contribution to the $\alpha$-R\'enyi entropy from a cubic trihedral corner in the boundary of the entangling region in $3+1$ dimensions for a massless free scalar.
The universal number, $v_{\alpha}$, is manifest as the coefficient of a scaling term that is logarithmic in the size of the entangling region.
Our numerical calculations find that this universal coefficient has both larger magnitude and the opposite sign to that induced by a smooth spherical entangling boundary in 
$3+1$ dimensions, for which there is a well-known subleading logarithmic scaling.  
Despite these differences, up to the uncertainty of our finite-size 
lattice calculations, the functional dependence of the trihedral coefficient $v_{\alpha}$ on the R\'enyi index $\alpha$ is indistinguishable from that for a sphere, which is known analytically for a massless free scalar.  
We comment on the possible source of this $\alpha$-dependence arising from the general structure of $(3+1)$-dimensional
conformal field theories, and suggest calculations past the free scalar which could further illuminate the general structure of the trihedral divergence
in the R\'enyi entropy.

\end{abstract}   
\maketitle 
\singlespacing 


\section{Introduction}  
\label{sec:intro}  

Bipartite entanglement entropies in quantum critical systems can harbour universal quantities that characterize the underlying theory.
Originating in the scaling dependence of entropies on the size of the entangled region, 
these quantities arise from different geometric features in the entangling boundary.
Such quantities provide both a new perspective on what constitutes a universal number as well as a concrete connection
between seemingly disparate physical theories arising in condensed matter, high energy field theory and gravity.
Amid the growing realization of their conceptual importance is the recognition that little is known about  
the number of undiscovered universal quantities, their numerical values,
and their relationship to conventional universality such as that arising from $n$-point correlation functions.
In spacetime dimensions higher than $d$=1+1, this uncertainty is present even for free theories, where the calculations
of entanglement entropies can be technically challenging.\cite{Casini1}  
Despite this challenge, the many different types of geometric features
available in higher-dimensional entangling surfaces offers a rich opportunity
to search for new universal quantities in the entanglement entropy.\cite{Casini3,Kallin2013,Bueno1,Bueno3,Witczak-Krempa_2016,Chojnacki_2016} 
In addition to giving information about features of the underlying critical theory, understanding these quantities in free theories
is a necessary precursor to their exploration in interacting critical points,\cite{Kallin2014,Fei2015}  such as those in real quantum materials or atomic systems.\cite{CuprateTc,BEC_Greiner,MagnetQCP1,MagnetQCP2}

\begin{figure}[ht]
        \centering
                \includegraphics[scale=0.52]{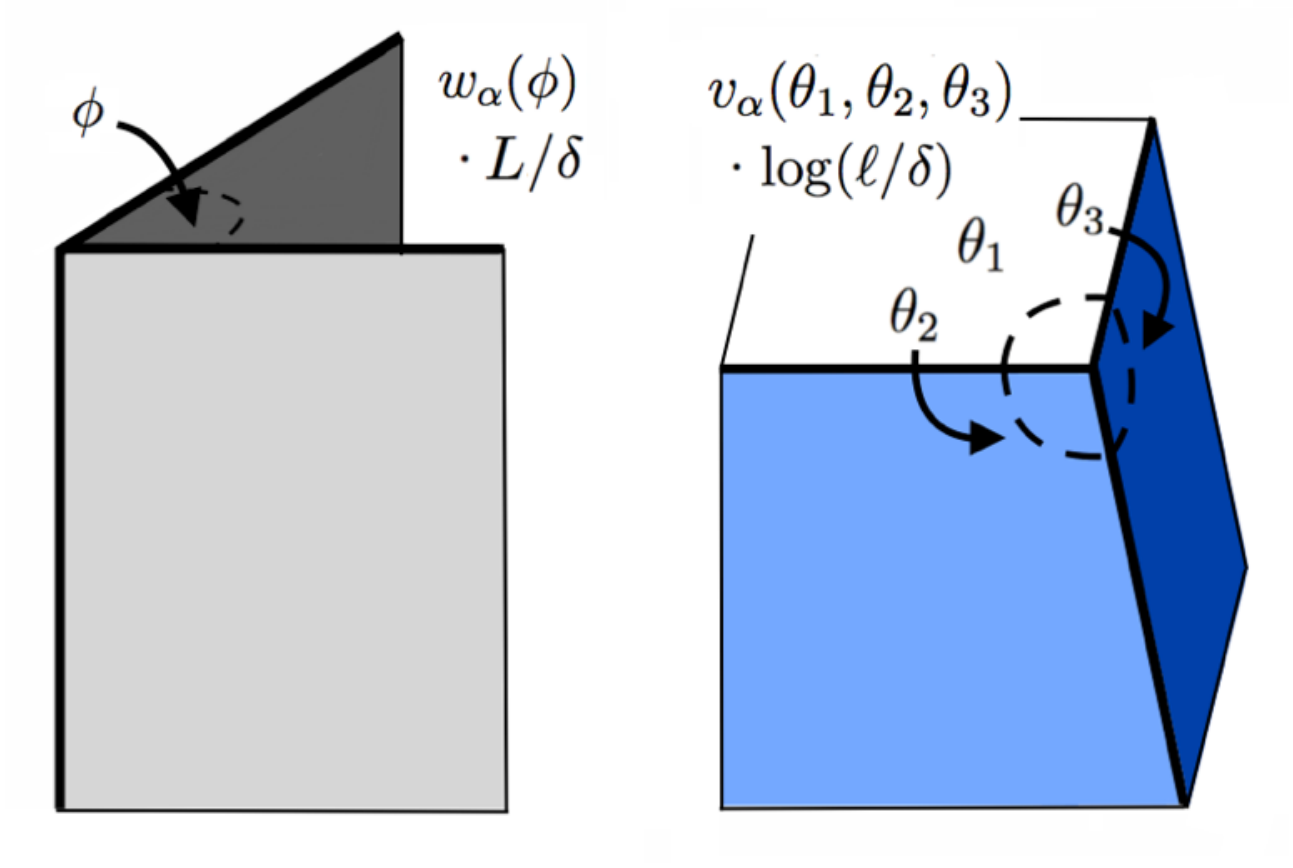}      
        \caption{Contributions to the \ren entropy of $(3+1)$-dimensional CFTs from (left) a wedge of opening angle $\phi$ (non-universal) and (right) a trihedral corner parametrized by angles $\theta_1$, $\theta_2$, $\theta_3$ (universal).    }
\labell{fig1}
\end{figure}

In this paper, we examine a particular universal quantity that arises in $d=3+1$ spacetime dimensions when the entangling geometry 
contains a (cubic) trihedral corner --- see Fig.~\ref{fig1}.  
This quantity $v_{\alpha}$ appears as the coefficient of a scaling term with a logarithmic dependence on the size of the entangling boundary. This coefficient has been computed previously in quantum Ising models at the infinite disorder \cite{Kovacs} and Wilson-Fisher fixed points.\cite{Devakul2014} 
The latter work observed that $v_{\alpha}$ exhibits a functional dependence on $\alpha$ that is close to that for the corresponding coefficient of  a spherical entangling surface.  
This observation suggests that it may be possible to understand the trihedral corner coefficient as coming from the singular limit of a smooth interface, \ie an eighth of that of the sphere. 
Here, we perform a numerical calculation of $v_{\alpha}$ for a massless free scalar field theory and show that while the functional dependence on $\alpha$
indeed closely matches that arising from a smooth sphere, in general both the magnitude and sign of the trihedral coefficient are different.
We suggest future calculations that may provide further insight into the universal content of the trihedral corner term.

\section{\Renyi entropies in $d=3+1$}
\label{sec:renn}

We consider the entanglement corresponding to spatial bipartitions of a physical system into a region $A$ and its complement $\bar A$, separated by a surface $\partial A$.
One can quantify the entanglement for such bipartite systems through the entanglement entropy $\see$ or, more generally, the \Renyi entropies $S_\al$.\cite{Renyi1,Renyi2} 
For our calculations, $A$ is bounded by a cubic trihedral corner, which is formed by three intersecting orthogonal planes.
The corresponding Hilbert space is bipartitioned such that $\mathcal{H}=\mathcal{H}_{A}\otimes \mathcal{H}_{\bar A}$ and then, given a state $\rho \in \mathcal{H}$, the $\alpha$-\Renyi entropy is defined as
\begin{equation}\label{renyi}
S_\alpha(A) = \frac{1}{1-\alpha} \ln \left[ \Tr \left( \rho_A^{\,\alpha} \right) \right] \, ,
\end{equation}
where $\rho_A = \Tr_{\bar A} \rho$ is the reduced density matrix associated with subregion $A$ and $\alpha$ is the \Renyi index. When Eq.~\req{renyi} can be evaluated for real values of $\alpha$, the entanglement entropy can in turn be obtained as $\lim_{\alpha\rightarrow 1} S_{\alpha}(A)=\see(A)=-\Tr \left( \rho_A \log \rho_A \right)$. 
We note that in many interacting systems, the \Renyi entropies can only be studied for positive integer $\alpha \ge 2$, which is accomplished by evaluating a
partition function on a multi-sheeted Riemann surface,\cite{Calabrese1,Calabrese2} as has recently been employed in quantum Monte Carlo simulations\cite{swap} and in experiments on ultra-cold atoms.\cite{Islam2015}

In the following, we begin with general scaling arguments that apply for all values of $\alpha$ for corner singularities in $d=3+1$. 
We then take another approach where we explore a smooth regularization of the cube. 
We comment on consistencies, inconsistencies and predictions related to these two approaches.

\subsection{General scaling arguments}

When computed for the vacuum state of a local Hamiltonian, \Renyi entropies are dominated by short-distance correlations across the entangling surface $\partial A$. 
These local correlations yield the celebrated ``area law" term as the leading contribution.\cite{sorkin,Bombelli1986,Srednicki1993} That is,
\beq
S_{\alpha}(A)=B_{\alpha}\,{\cal A}/\delta^{d-2}+\cdots\,,
\label{got}
\eeq
where we are considering a quantum field theory living in $d$ spacetime dimensions.
In this expression, the coefficient $B_{\alpha}$ is non-universal, \ie regulator dependent, such that it depends on the procedure used to regulate the calculation.
$\cal A$ is the area of the entangling surface $\partial A$ and $\delta$ is a short-distance cutoff, \eg the lattice spacing. 
In general, the subleading contributions indicated by the ellipsis in Eq.~\reef{got} include further power-law divergences and the corresponding coefficients also depend on the details of the regulator. 
However, a regulator-independent coefficient providing well-defined information about the underlying theory appears if there is a subleading contribution that scales logarithmically with $\ell/\delta$, where $\ell$ is a length scale characteristic of the size of the region $A$. 
For example, in $d=2+1$, introducing a sharp corner in the entangling surface produces such a logarithmic contribution $S^{\rm univ}_{\alpha,\rm corner}=-a_{\alpha}(\theta)\log(\ell/\delta)$, where the universal coefficient $a_{\alpha}(\theta)$ is a function of the opening angle $\theta$ of the corner and the \Renyi index $\alpha$.\cite{Casini1,Casini2,Casini3,Hirata,Myers:2012vs,Bueno1,Bueno2,Bueno3,Bueno5,Miao2015,Elvang,Kallin2013,Stoudenmire2014,helmes14,devakul14,Kallin2014,Sahoo2016,Helmes2016,DeNobili:2016nmj}
In $d=3+1$, geometric singularities in $\partial A$ can also produce similar universal contributions in $S_{\alpha}(A)$.\cite{Klebanov:2012yf,Myers:2012vs,Bueno4,Devakul2014,Kovacs}
 
For the present discussion, let us focus on a three-dimensional region $A$ which is a polyhedron auch that the entangling surface $\partial A$ consists of flat polygonal faces, straight edges and sharp corners or vertices. We limit our discussion to the \Renyi entropies in a conformal field theory (CFT), in which case only geometric scales appear in $S_\alpha$.\footnote{In a more general QFT, the dimensionless  coefficients $B_\al$, $w_\al$ and $v_\al$ may also depend on the combinations $\mu_i\delta$, where the $\mu_i$ denote various mass scales appearing in the QFT. See, for example, discussion in Ref.~\onlinecite{Myers:2012vs}.} For such a geometry, $S_\alpha(A)$ are expected to scale such that
 \beqa
S_{\alpha, \rm poly} &=& B_\alpha \, {\cal A}/{\delta^2} + \sum_i w_\al(\phi_i) \, {{L_i}}/{\delta} \label{eq:Spoly} \\
&&{}+ \sum_i v_\alpha(\theta_{i,1},\theta_{i,2},\theta_{i,3})\log\!\left(\ell/\delta \right) + \mathcal{O}((\ell/\delta)^0) \, .
\nonumber
\eeqa
Beyond the area law term, the first subleading contribution arises from the edges on the boundary of the polyhedron.\cite{Myers:2012vs,Klebanov:2012yf} Here $L_i$ is the length of the $i^{\text{th}}$ edge.\footnote{We examine the functional form of this edge contribution to the \Renyi entropy in more detail in Appendix~\ref{app_edge}.}  The (non-universal) coefficients $w_\al(\phi_i)$ depend on the opening angle $\phi_i$ between the two faces intersecting at the edge. The logarithmic term in Eq.~\reef{eq:Spoly} arises from the vertices in $\partial A$.
\footnote{In this term, $\ell$ refers to some length scale characteristic of the geometry of region $A$. Note that for the cube in Eq.~\reef{eq:S_octant}, there is a single scale $L$ that defines the area and the edge lengths. In this case, $L$ naturally appears in the logarithmic term.}  
As shown in Fig.~\ref{fig1}, the universal coefficient $v_\alpha(\theta_{i,1},\theta_{i,2},\theta_{i,3})$ at the $i^{\text{th}}$ vertex is a function of the angles $\theta_{i,1}$, $\theta_{i,2}$ and $\theta_{i,3}$ between each of the three pairs of (adjacent) edges that end at the vertex. 
For the entanglement entropy, $w_1(\phi)\le 0$ follows from strong subadditivity;\footnote{This inequality follows from a simple generalization of the argument in Ref.~\onlinecite{Casini2} to higher dimensions.} 
however, there are no known analogous arguments that rigorously fix the sign of either $w_\al$ for general $\al$ or of $v_\al$. However, intuitive arguments can be made to suggest that $w_\al\le0$ and $v_\al\ge0$.\footnote{For example, we might imagine that the area-law term simply counts Bell-pair correlations across the smooth faces in $\partial A$. Then $w_\al\le0$ follows since the edge terms must compensate for over-counting the correlations of the degrees of freedom near the edges.} These signs were confirmed for the explicit examples in Refs.~\onlinecite{Devakul2014,Kovacs} and are reproduced in our calculations below.

This paper focusses on the universal coefficient $v_\al$ of the logarithmic contribution to $S_\alpha$
arising from a trihedral corner (formed by three intersecting faces as in Fig.~\ref{fig1}).  For simplicity, we examine the cubic case where the opening angles are all $\pi/2$ and thus henceforth we omit from our notation the angular dependence of $v_\al$. 
For a cube of  dimension $L$, Eq.~\reef{eq:Spoly} then simplifies to 
 \begin{equation}
S_{\alpha, \rm cube} = B_\alpha \, \frac{6L^2}{\delta^2} + w_\al  \frac{12L}{\delta}  + 8v_\alpha\log\!\left(L/\delta \right) + \mathcal{O}((L/\delta)^0) \,.
\label{eq:S_octant}
\end{equation}
We are particularly interested in the dependence of the universal corner coefficient $v_\al$ on the \Renyi index $\al$.

\subsection{Smooth entangling surfaces}
\label{sec:smoothSurfaces}

As remarked in the introduction, one might hope to understand the trihedral corner coefficient as coming from the singular limit of a smooth entangling surface. Hence,
in this section, we review the structure of universal contributions to the \ren entropy for smooth entangling surfaces in $(3+1)$-dimensional CFTs. In this case,\footnote{Implicitly, we are considering the region $A$ to lie on a constant time slice in flat space with $d=4$.} the \ren entropy takes the form
 \beq
S_{\alpha, \rm smooth} = B_\alpha \, {\cal A}/{\delta^2} +  u_\alpha \log\!\left(\ell/\delta \right) + \mathcal{O}((\ell/\delta)^0) \, ,
\label{smooth}
\eeq
where the universal coefficient $u_{\al}$ is determined by the geometry of the entangling surface according to\cite{Fursaev:2012mp,Lee:2014xwa}
\begin{align}
u_{\al}=-\int_{\partial A}  \frac{d^2y\sqrt{h}}{2\pi} \,\left[ f_a(\al)\,\mathcal{R}+f_{b}(\al)\,\hat{K}^2\right]\, .
\labell{gamer}
\end{align}
In this expression, $h=\det(h_{ij})$ is the determinant of the induced metric  on $\partial A$, $\mathcal{R}$ is the Ricci scalar of this metric, and $\hat{K}^2\equiv K_{ij}^{a}K_{ji}^{a}- (h^{ij}K^{a}_{ji})^2/2$ where $K_{ij}^{a}$ is the extrinsic curvature of the entangling surface.

The two coefficients $f_a(\al)$ and $f_b(\al)$ in Eq.~\req{gamer} contain universal information that characterizes the underlying CFT. 
In particular, in the limit $\al\to1$, these functions yield the central charges in the trace anomaly,\cite{Solodukhin:2008dh} \ie $f_a\to a$ and $f_b\to c$. 
In the case of a massless free scalar, which we study here, these functions are\cite{Fursaev:2012mp,Lee:2014xwa}$^,$\footnote{The result for $f^\mt{scalar}_b(\al)$ in Eq.~\reef{name} relies on the equality $f_b(\al)=f_c(\al)$ where $f_c(\al)$ is a third universal coefficient in Eq.~\reef{gamer} proportional to the Weyl curvature of the background spacetime. This equality is known {\it not} to hold for general CFTs,\cite{dong1,dong2} but Ref.~\onlinecite{Lee:2014xwa} provided strong numerical evidence of its  validity for the free scalar. Recently, Ref.~\onlinecite{displace} also provided an analytic proof that $f^\mt{scalar}_b(\al)=f^\mt{scalar}_c(\al)$\label{fuo}} 
\begin{equation}\label{name}
f^\mt{scalar}_a(\al)=\frac{1}{3}f^\mt{scalar}_b(\al)=\frac{(\al+1)(\al^2+1)}{1440\al^3}\, .
\end{equation}

We note that the first contribution in Eq.~\eqref{gamer} is topological such that the integral multiplying $f_a(\al)$ yields twice the Euler characteristic $\chi$ of the entangling surface. 
In particular, this contribution is the same for a sphere $S^2$ and a cube $C^2$, \ie $\chi(S^2) = \chi(C^2) = 2$. 
For a sphere, this term represents the only contribution to the universal part of the \ren entropy since $\hat{K}^2(S^2)=0$ and so one finds
\beq
S^{ \rm univ}_\al(S^2)= -4\,f_a(\al) \log\left(R/\delta\right)\, ,
\labell{game2d3}
\eeq
where $R$ is the radius of the sphere. When the entangling surface is a cube, the curvature is entirely concentrated in the eight corners and hence,
na\"ively, one might set $v_\al=-f_a(\al)/2$ in Eq.~\eqref{eq:S_octant} for the universal contribution from each individual corner.  We examine this idea in more detail below.

\subsection{Smoothed cube}
\label{Scube}

\begin{figure}[ht]
        \centering
                \includegraphics[scale=0.4]{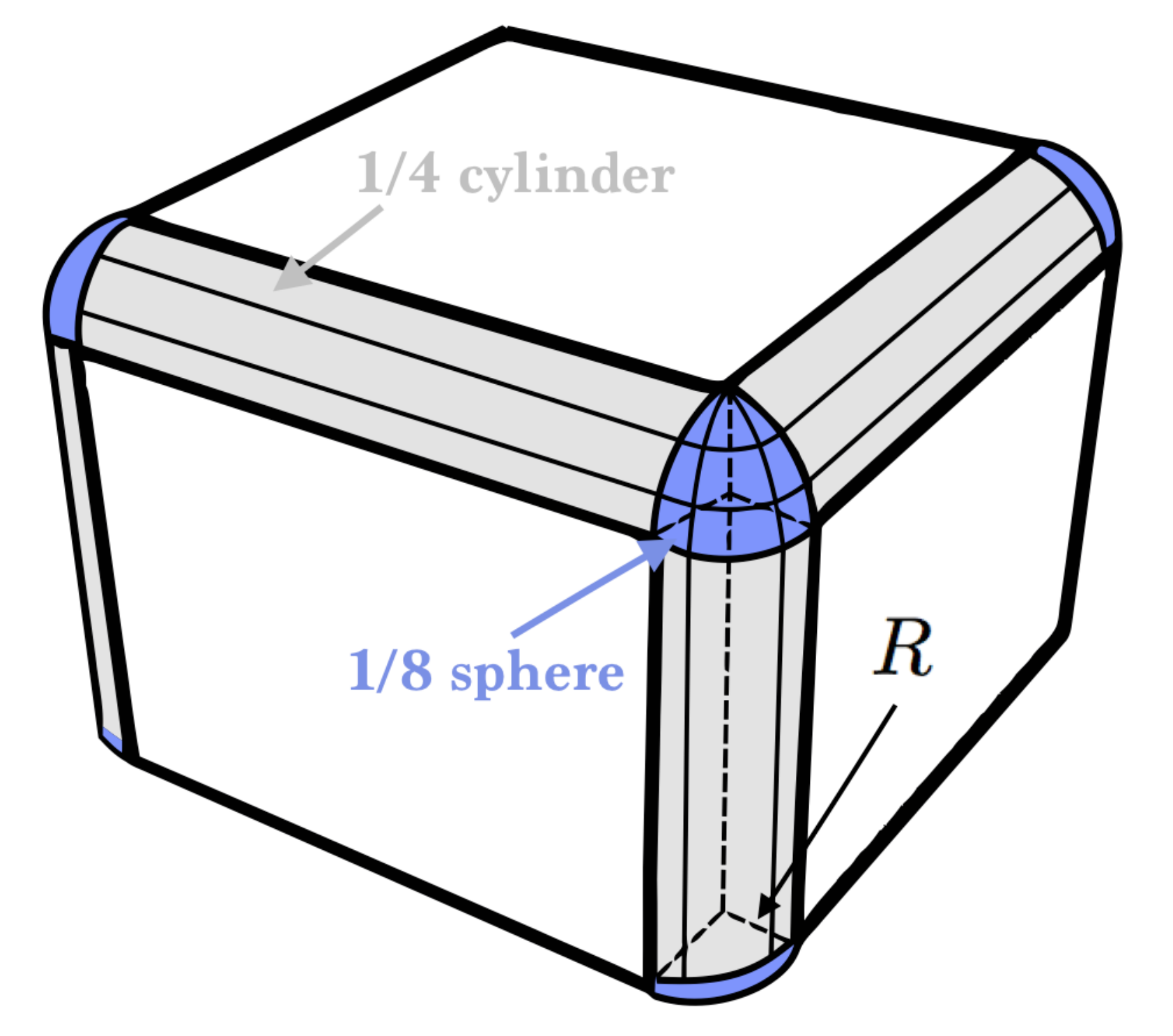}      
        \caption{Rounded cube ${\widetilde C}^2$. 
        The corners and edges are respectively replaced by eighths of a sphere and quarters of a cylinder of radius $R=\veps L$ with $\veps\ll1$. 
}
\labell{fig2}
\end{figure} 

In Ref.~\onlinecite{Devakul2014}, a numerical estimation of the coefficient $v_\al$ for a cubic trihedral corner was obtained for the Ising theory, with a magnitude very close to $1/8$ of the value of the universal coefficient appearing in Eq.~\reef{game2d3}, \ie  $v_\al\simeq f_a(\al)/2$. 
We comment that this result has the wrong sign compared to $S^{ \rm univ}_\al(S^2)$, but since this difference was not noticed at the time,
the result in Ref.~\onlinecite{Devakul2014} led to the suggestion that the trihedral corner coefficient might be understood as coming 
from the singular limit of a smooth entangling surface.
In order to explore this,
we consider a smoothed cube in which the edges and vertices have been rounded off and replaced by cylinders and spheres, respectively.
In particular, as shown in Fig.~\ref{fig2}, we consider a rounded cube ${\widetilde C}^2$ where each of the eight corners is replaced by an eighth of a sphere of radius $R=\veps L$ with $\veps\ll1$. Further, each of the twelve edges is replaced by a quarter cylinder of  radius $R=\veps L$ and length $L\,(1-2\gamma_0 \veps)$ where $\gamma_0$ is a fixed constant of $\cO(1)$. A natural choice for this constant would correspond to $\gamma_0=1$, which fixes the central width of ${\widetilde C}^2$ to be $L$ for all values of $\veps$.\footnote{One can also consider more elaborate schemes. For example, one can choose $\gamma_0$ to  fix the area of ${\widetilde C}^2$ such that it matches that of $C^2$ instead, \ie one can fix $\cA= 6L^2$ in Eq.~\reef{area}.
Observe that, to within the resolution of the short-distance cutoff $\delta$, any of these choices yields the same cube $C^2$ in the limit $R\to\gamma_1\delta$.}
By design, in the limit $\veps\to0$ we recover the usual cube with sharp edges and corners. Hence, it does not seem \emph{a priori} unreasonable to expect that the trihedral corner coefficient can be extracted from the resulting \ren entropy, however, as we show below, the situation is more subtle.

Applying Eqs.~\reef{smooth} and \req{gamer}, the \ren entropy of our smoothed cube becomes
 \beqa
S_{\alpha}({\widetilde C}^2) &=& B_\alpha \, {\cal A}/{\delta^2} -\left[4f_a(\al)+\left(\frac{3}{2\veps}-3\gamma_0\right) f_b(\al)\right]
\nonumber\\
&&\qquad\qquad\times\ \log\!\left(L/\delta \right) + \mathcal{O}((L/\delta)^0)\,,
\label{game1hh}
\eeqa
where the area is given by
\beq
\cA=6L^2+L^2\veps\,(6\pi-24\gamma_0)+L^2\veps^2 (4\pi-12\pi\gamma_0+24\gamma_0^2) \, .
\label{area}
\eeq
We can see that the limit $\veps\to0$ is problematic in Eq.~\reef{game1hh} since the coefficient of the logarithmic term diverges.  To produce regulated (\ie finite) \ren entropies, we take the limit $\veps \to \gamma_1\delta/L$ or $R\rightarrow \gamma_1\delta$ where $\gamma_1$ is again some $\cO(1)$ constant. In this limit, the edges and corners of the cube are still rounded at a scale of the order of the short-distance cutoff $\delta$. In particular, 
Eq.~\reef{game1hh} yields
 \beqa
S_{\alpha}({\widetilde C}^2) &=& B_\alpha \, \frac{6L^2}{\delta^2} -\frac{3f_{b}(\alpha)}{2\gamma_1}\,\frac{L}\delta\log(L/\delta)
\label{large}\\
&&\quad + w_\al  \frac{12L}{\delta}  + 8v_\alpha\log\!\left(L/\delta \right) + \mathcal{O}((L/\delta)^0) \,,
\nonumber
\eeqa
where $w_\al=-B_{\alpha}\gamma_1(2\gamma_0-\pi/2)$ and
\beq
v_\al=-\frac12 f_a(\al)+\frac38\gamma_0\, f_b(\al)\,.
\label{larger}
\eeq
This result is problematic in two ways: First, we observe the appearance of a new divergence of the form $L/\delta\,\log(L/\delta)$, which is incompatible with the form expected in Eq.~\eqref{eq:S_octant} (see Appendix~\ref{app_edge}). Second, the coefficient of the logarithmic term is ambiguous because of the appearance of $\gamma_0$ in Eq.~\reef{larger}, \ie our desired ``universal'' coefficient depends on the details of the regulator.

A possible resolution of both of these problems is that Eq.~\reef{game1hh}, or Eq.~\reef{smooth}, simply does not describe the \Renyi entropies with sufficient accuracy to take the desired limit. That is, originally $\veps$ is small but independent of the ratio $\delta/L$, and hence any terms of the form $\log\veps$ or $\veps^{-1}\log\veps$ are concealed in the $\mathcal{O}((L/\delta)^0)$ contributions. However, with the limit $\veps \to \gamma_1\delta/L$, such terms emerge to modify the terms explicitly enumerated in Eq.~\reef{large}. 
That is, order-one contributions may be building up in the singular limit to restore the universality of the logarithmic coefficient and to cancel the unanticipated $L/\delta\,\log( L/\delta)$ contribution. We expand on this suggestion in the discussion in Section~\ref{discuss}. 

Regardless, our smoothed-cube calculation suggests that the universal coefficient $v_\al$ is some linear combination of $f_a(\al)$ and $f_b(\al)$.  
Further, 
Eq.~\eqref{name} shows that the massless free scalar is a special case where both of these coefficients have the same dependence on the \ren index. 
Hence, independent of the precise linear combination, the above calculation suggests that 
\beq
\frac{v^{\mt{scalar}}_\al}{v^{\mt{scalar}}_1}= \frac{(\al+1)(\al^2+1)}{4\,\al^3}\, .
\labell{game1}
\eeq
Below, we compare this prediction with the $\al$-dependence calculated numerically for the free scalar, and show that we find good agreement to within numerical accuracy.  Note that this argument predicts that other theories 
in general have a different dependence on $\alpha$, depending on the precise linear combination of $f_a(\al)$ and $f_b(\al)$ appearing in $v_\al$. 

\section{Numerical methods for the free scalar field}
\label{sec:Methods}
In this section we perform direct calculations of the \Renyi entropies $S_\alpha(A)$ for a free real scalar field on a three-dimensional simple cubic lattice.
At each lattice site $i$, there exists a bosonic field $\phi_i$ and its conjugate momentum $\pi_i$, whose dynamics are controlled by the Hamiltonian
\begin{equation}
H = 
\frac{1}{2}\sum_{i=1}^N \left( \pi_i^2 + m^2\phi_i^2 \right) 
+ \frac{1}{2} \sum_{\langle i j \rangle} \left( \phi_j - \phi_i  \right)^2\, .
\label{eq:Ham_free}
\end{equation}
In this expression, the first sum is over the $N$ lattice sites, while the second is over all nearest-neighbour pairs of sites. $m$ is the mass of the scalar field.
For a lattice in three spatial dimensions, the total number of lattice sites is $N = L_x L_y L_z$, where $L_x$, $L_y$ and $L_z$ are the linear lattice dimensions along $x$, $y$ and $z$ respectively.

This Gaussian theory has the appealing property that $S_\alpha(A)$ can be obtained for any $\alpha$ from knowledge of the two-point functions of $\phi_i$ and $\pi_i$ at lattice points within region $A$ --- see Section~\ref{sec:PeschelTrick}. 
In order to isolate the logarithmic corner contribution, we use these calculations along with techniques from the numerical linked-cluster expansion\cite{Rigol2006,Rigol2007_1,Rigol2007_2,Tang2013} (NLCE), as described in Section~\ref{sec:NLCE}.

We focus on the case where the boson is massless ($m=0$) such that the energy spectrum is gapless and we have a scale-invariant critical theory.
However, the techniques described in this section are equally applicable to free bosons of any mass. Further, note that in the following we measure lengths in units of the lattice spacing, which we set to unity for simplicity, \ie $\delta=1$.

\subsection{\Renyi entropies for free scalars}
\label{sec:PeschelTrick}
Here we explain how \ren entropies for free scalars can be computed from the correlators $\left\langle \phi_i \phi_j \right\rangle$ and $\left\langle \pi_i \pi_j \right\rangle$ as first introduced in Ref.~\onlinecite{Peschel2003}.
The Hamiltonian in Eq.~\eqref{eq:Ham_free} can be written in the more general quadratic (Gaussian) form
\begin{equation}
H = 
\frac{1}{2}\sum_{i=1}^N \pi_i^2
+ \frac{1}{2} \sum_{ i j } \phi_i M_{ij} \phi_j  ,
\label{eq:Ham_quadratic}
\end{equation}
where $M$ is an $N \times N$ matrix that takes into account the boundary conditions of the finite lattice.
The groundstate two-point correlators are given in terms of this matrix as
\begin{align}
X_{ij} &\equiv \left\langle \phi_i \phi_j \right\rangle = \frac{1}{2} \left( M^{-1/2} \right)_{ij} \label{eq:XP}\, , \\
P_{ij} &\equiv \left\langle \pi_i \pi_j \right\rangle = \frac{1}{2} \left( M^{1/2} \right)_{ij}\,  . \nonumber
\end{align}
In order to calculate the \Renyi entropies corresponding to a given region $A$ for a quadratic Hamiltonian, one only needs to calculate $X_{ij}$ and $P_{ij}$ for pairs of sites $i,j \in A$.\cite{Peschel2003}
These region-restricted correlation functions define the matrix $C_A \equiv \sqrt{X_A P_A}$, where $X_A$ and $P_A$ are defined as in Eq.~\eqref{eq:XP} but with $i$ and $j$ labelling the $N_A$ sites in $A$.
The von Neumann and \Renyi entropies are then given in terms of the eigenvalues $\nu_k$ of $C_A$ as\cite{Casini2009}
\begin{align}
S_1 (A) = 
\sum_{k=1}^{N_A} \bigg[ & \left( \nu_k + \frac{1}{2} \right) \log \left( \nu_k + \frac{1}{2} \right) \\ \nonumber
&- \left( \nu_k - \frac{1}{2} \right) \log \left( \nu_k - \frac{1}{2} \right) \bigg]\, , \label{eq:vonNeumann_EE_eigvals}
\end{align}
and
\begin{equation}
S_\alpha(A) = 
\frac{1}{\alpha-1} \sum_{k=1}^{N_A}
\log \left[ \left( \nu_k + \frac{1}{2} \right)^\alpha - \left( \nu_k - \frac{1}{2} \right)^\alpha \right]\, .\label{eq:Renyi_EE_eigvals}
\end{equation}

We note that the $\left\langle \phi_i \phi_j \right\rangle$ correlators --- and consequently the above entropies --- diverge in the case where the boson is massless ($m=0$) and the lattice has periodic boundary conditions (PBC) in all lattice directions.
Note however that since we are using these expressions as the ``cluster solver'' for the NLCE procedure, 
which requires lattice clusters with open boundary conditions, these divergences do not pose a threat.

\subsection{Numerical linked-cluster expansion}
\label{sec:NLCE}

The NLCE is a powerful method that combines measurements of a suitable property on various finite-sized lattice clusters to obtain a sequence of approximations for the corresponding property in the thermodynamic limit $L \to \infty$. 
At a given length scale or ``order" $\ell$, this numerical expansion uses sums and differences of finite clusters to systematically cancel off lower-order finite-size and boundary effects.
As a result, at a given order this procedure is capable of accessing longer-range correlations than direct calculations on finite toroidal systems of the same size. 
This feature becomes especially advantageous when studying behavior at a critical point where the correlation length diverges.

For our present purposes, the NLCE offers the additional advantage that it can perform calculations in such a way as to isolate the corner contribution to the \Renyi entopies $v_{\alpha}\log(\ell/\delta)$ from both the edge contributions $w_\al\cdot (\ell/\delta)$ and the leading area law in Eq.~\eqref{eq:Spoly}.
In this section, we discuss general properties of the NLCE as well as the techniques necessary to isolate this three-dimensional corner contribution at each cluster order.
Analogous isolation techniques have been used with success to study the corner coefficient for various $(2+1)$-dimensional critical systems.\cite{Kallin2013,Kallin2014,Stoudenmire2014,Sahoo2016,Helmes2016}

At the most general level, the NLCE method can be used to study any property $\mathcal{P}$ that is
well defined in the thermodynamic limit (such as an extensive or an intensive property).
The NLCE calculates $\mathcal{P}$ for a lattice system $\mathcal{L}$ by summing 
contributions from individual clusters according to
\begin{equation}
\mathcal{P}\left( \mathcal{L} \right) = \sum_{c} W(c)\,  ,
\label{eq:NLCE_property}
\end{equation}
where the sum is over all clusters that can be embedded in the lattice
and $W(c)$ is the weight of the cluster.  
This weight is defined recursively as
\begin{equation}
W(c) = \mathcal{P}(c) - \sum_{s \in c} W(s)\, ,
\label{eq:NLCE_weight}
\end{equation} 
where the sum is over all subclusters $s$ contained in $c$.
This subgraph subtraction procedure elimintates from $W(c)$ the contributions to $\mathcal{P}(c)$ that have already been accounted for in the smaller subclusters.
Since we consider properties that are suitably normalized and have a well-defined thermodynamic limit (\ie extensive
or intensive properties), only connected (or linked) clusters have non-zero weight. 
Hence the sums in Eqs.~\eqref{eq:NLCE_property} and~\eqref{eq:NLCE_weight} can be restricted to all linked clusters. 

\begin{figure*}[ht]
	\begin{center}
	\def\svgwidth{\textwidth} 
	\input{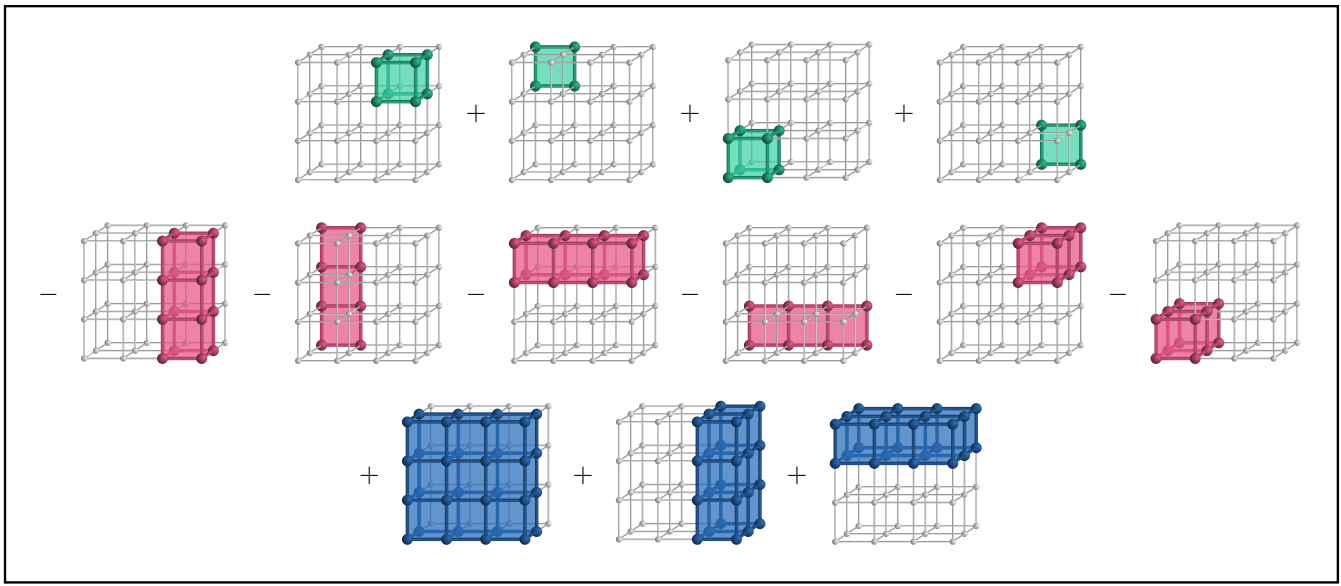}
	\end{center}
	\caption{The subtraction procedure used to calculate $\mathcal{P}_r(c)$ for the $3 \times 4 \times 4$ cluster $c$ and for a given vertex location $r$ (not pictured). 
	We add the values of $S_\alpha(A)$ corresponding to the four octants in the top row, subtract the values for the six quadrants in the middle row, and add the values for the three half-planes in the bottom row. We divide the resulting sum by four in order to obtain $\mathcal{P}_r(c)$.  	}
	\label{fig:cuts}
\end{figure*} 

In a translationally invariant system, all clusters that are related by translations have the same weights and make identical contributions to an extensive property $\mathcal{P}$.
For a given cluster, there are $N$ such clusters $c$ with the same weight such that the expression for $\mathcal{P}$ reduces to
\begin{equation}
\mathcal{P}\left( \mathcal{L} \right)/N = \sum_{c'} W(c') \, ,
\label{eq:NLCE_property2}
\end{equation}
where the clusters $c'$ are defined modulo translation. 

One can often further reduce the number of clusters required in the above sums by exploiting other symmetries of the lattice system.
When the weights are the same for a given class of clusters, then one can write
\begin{equation}
\mathcal{P}\left( \mathcal{L} \right)/N = \sum_{c''} \mathbb{L}(c'') \times W(c'')  \,,
\label{eq:NLCE_property3}
\end{equation}
where the sum is now over representative clusters $c''$ from each cluster class. 
The quantity $\mathbb{L}(c'')$, called the lattice constant or the embedding factor of the cluster $c''$,
is the number of ways per lattice site that a cluster of class $c''$ can be embedded in the lattice.

So far the graphical basis for the NLCE is the same as for a series expansion in some variable, 
such as the inverse temperature $\beta$ in the high temperature series expansion (HTSE). 
However, unlike the HTSE where the goal is to maximize the order up to which the expansions are performed, 
here the goal is to include contributions from representative clusters of maximal size. 
Since the number of possible clusters grows rapidly with order, it is useful to further restrict the types of clusters considered in the expansions.\cite{Kallin2013}

Specifically, in $D$ spatial dimensions, every cluster can be uniquely associated with a $D$-dimensional rectangle (called a cuboid in 
3D),
defined as the smallest volume in which the cluster can be fully embedded. 
Thus, one can limit the calculations to cuboidal clusters only. 
In three spatial dimension, one can then use Eq.~\eqref{eq:NLCE_property3}
with the sum restricted to regular $u_x \times u_y \times u_z$ cuboids with 6 faces, 8 vertices and 12 edges each. 
Here $u_x$, $u_y$ and $u_z$ are integer lengths measured in units of the lattice spacing.
The enumeration of such clusters is trivial since their count (or embedding factor) $\mathbb{L}(c'')=1, 3 \text{ or } 6$ just depends on
the symmetry of the cuboid. 
The subgraph counts of smaller cuboids in larger ones (needed in Eq.~\eqref{eq:NLCE_weight}) are also trivial. Because of the unique
association of each cluster with a cuboid the subgraph subtraction scheme works using cuboids only and thus the entire
computational burden is on calculating properties of finite cuboids. 
Note that clusters in NLCE correspond to actual clusters
that can be embedded in the infinite lattice and therefore there is no periodic boundary condition on the finite clusters. 

So far our discussion has focused on extensive properties, which get equal contributions from every region of the lattice.
However, the NLCE method applies equally well to an intensive property $\mathcal{P_\text{int}}$ that is defined with respect to some particular location in the lattice.
In this case, the clusters must be rooted with respect to
the special location where the property is defined.
As a result, the factor of $N$ that led to the simplification of Eq.~\eqref{eq:NLCE_property2} is no longer present.
However, one can create a corresponding extensive quantity by moving the localized quantity to different sites of the lattice and adding up the contributions for different locations.
In a translationally invariant system, the simplying factor of $N$ is then restored and such contributions are all the same so that $\mathcal{P_\text{ext}} = N \mathcal{P_\text{int}}$.
Then the left-hand side of Eqs.~\eqref{eq:NLCE_property2} and~\eqref{eq:NLCE_property3} is $\mathcal{P_\text{ext}} / N = \mathcal{P_\text{int}}$.

Similar to previous calculations in lower dimensions,\cite{Kallin2013,Kallin2014,Stoudenmire2014,Sahoo2016,Helmes2016} we define the desired intensive property $\mathcal{P}$ to be the isolated trihedral vertex contribution to the entanglement entropy $S_\alpha$.
We conceptually imagine a single vertex of interest as arising from an octant of the solid geometry in three spatial dimensions.
Then, each cuboidal cluster $c$ arising in the NLCE calculation is embedded in all possible ways around this vertex in order to create a corresponding extensive quantity as described above.
All possible rotations of each cluster are accounted for by $\mathbb{L}(c)$ in Eq~\eqref{eq:NLCE_property3}.
For a given cuboidal cluster, there are $(u_x -1)( u_y -1)( u_z - 1)$ possible locations $r$ for this vertex within 
the cluster, which can be alternatively thought of as translating the cluster with respect to a fixed location of the
corner. 
The overall contribution from the cluster is obtained by summing over all of these possible locations such that $N\mathcal{P}(c) = \sum_r \mathcal{P}_r(c)$.

To proceed with the calculation of the entanglement properties, we construct our clusters and subclusters to be regular 
$u_x \times u_y \times u_z$ cuboids as described above. 
For the corner entanglement entropy, there is no contribution from clusters
where any of $u_x$, $u_y$ or $u_z$ take value one. 
We define the length scale (order) of a given cluster to be the maximum of these linear dimensions such that $\ell = \max\{u_x,u_y,u_z\}$.
Each cluster imposes Dirichlet open boundary conditions, with the field $\phi$ constrained to be zero for all lattice sites outside of the cluster.

Finally, in order to isolate the subleading trihedral vertex contribution to the \Renyi entropies, we perform a cluster-by-cluster subtraction procedure. 
For each vertex location $r$ within the cluster $c$, we combine the values of $S_\alpha(A)$ corresponding to 13 different bipartitions $\{A, \overline{A}\}$ of the cluster as illustrated in Fig.~\ref{fig:cuts}.
This combination of \Renyi entopies allows us to intrinsically cancel the leading contributions from the area law (which are proportional to $B_\alpha \, {\ell^2}$) and the 90-degree edges (proportional to $ w_\al \, {\ell}$) such that $\mathcal{P}_r(c)$ (and, in turn, $\mathcal{P}(c)$) corresponds to 
only the trihedral corner contribution to the entropy.
In practice, we take advantage of the symmetries present in the system in order to reduce the number of cluster bipartitions from 13 to 7.
The correlation function methods described in Section~\ref{sec:PeschelTrick} act as our cluster solver such that these methods are used to calculate all needed \Renyi entropies in the above sums for each cluster.

\section{Results}
\label{sec:Results}
In this section, we use the methods outlined in Section~\ref{sec:Methods} to calculate the trihedral corner coefficient in the \Renyi entropies of a massless free scalar in $(3+1)$-dimensions.
Using the NLCE procedure described in Section~\ref{sec:NLCE}, we isolate the corner contribution $\mathcal{P}_\alpha(\ell)$ to the \Renyi entropy $S_\alpha$ by performing calculations on clusters up to order $\ell$ (the maximum linear dimension of a given cluster).
In this section we examine the behavior of $\mathcal{P}_\alpha(\ell)$ as a function of $\ell$ with the goal of studying the vertex coefficient $v_{\alpha}$. 
From Eq.~\eqref{eq:Spoly}, we expect for a single vertex,
\begin{equation}
\mathcal{P}_\alpha = v_{\alpha} \log \ell + d_\alpha + \ldots,
\label{eq:P_alpha_scaling}
\end{equation}
where $d_\alpha$ is a subleading constant and the ellipsis indicates additional (unknown) subleading terms that should vanish as $\ell \to \infty$ (or $\delta\to0$). Recall that $\ell$ is measured in units of the lattice spacing.

\begin{figure}[t]
	\begin{center}
	\includegraphics{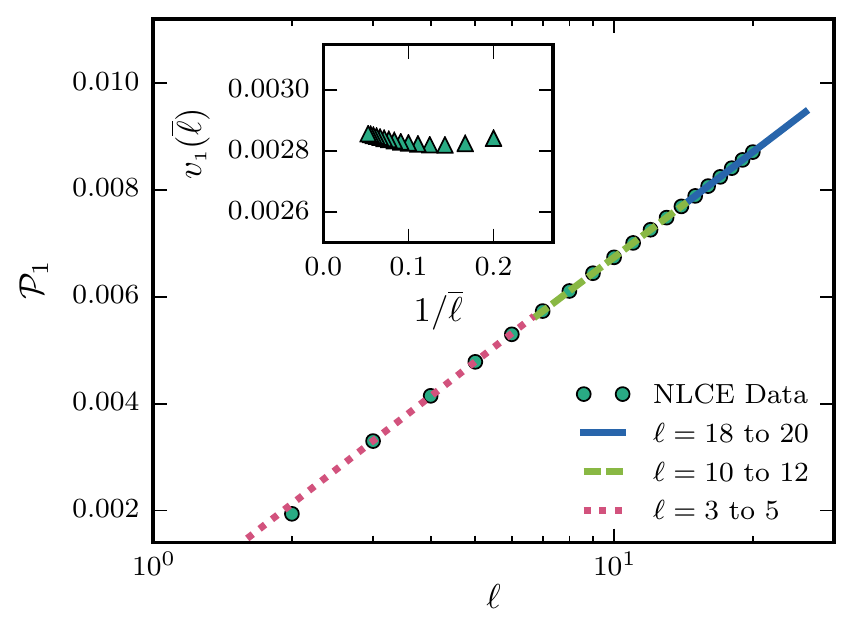}
	\includegraphics{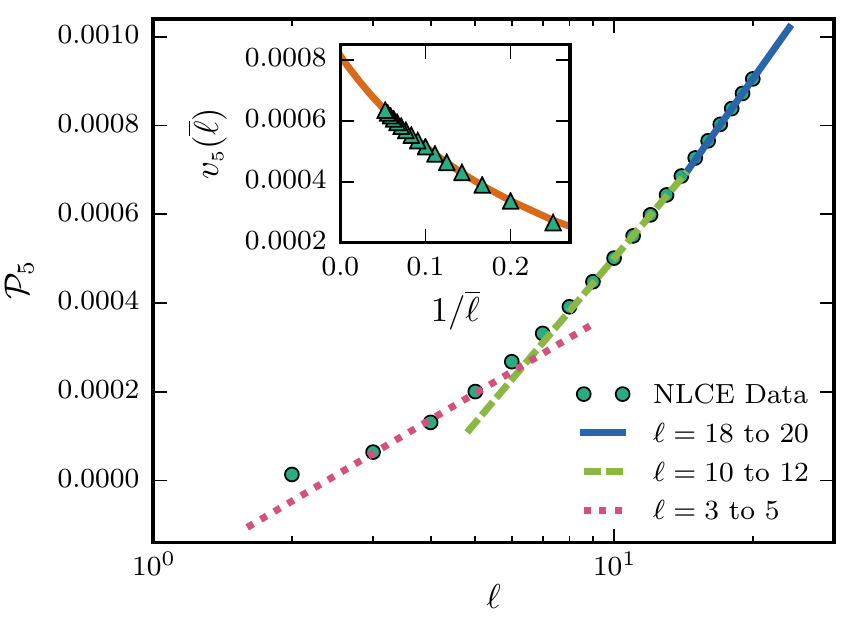}
	\end{center}
	\caption{Fits of the corner contribution $\mathcal{P}_\alpha$ to the equation $v_{\alpha} \log \ell + d_\alpha$ for \Renyi indices $\alpha=1$ (top) and $\alpha=5$ (bottom). 
	The insets illustrate how the coefficients $v_{\alpha}$ extracted from these fits depend on the range of values of $\ell$.
	For $\alpha>1$, we perform a second fit to extrapolate to the thermodynamic limit, as explained in the main text.
	}
	\label{fig:VAlpha_fits}
\end{figure}

\begin{figure*}[tp]
	\begin{center}
	\includegraphics{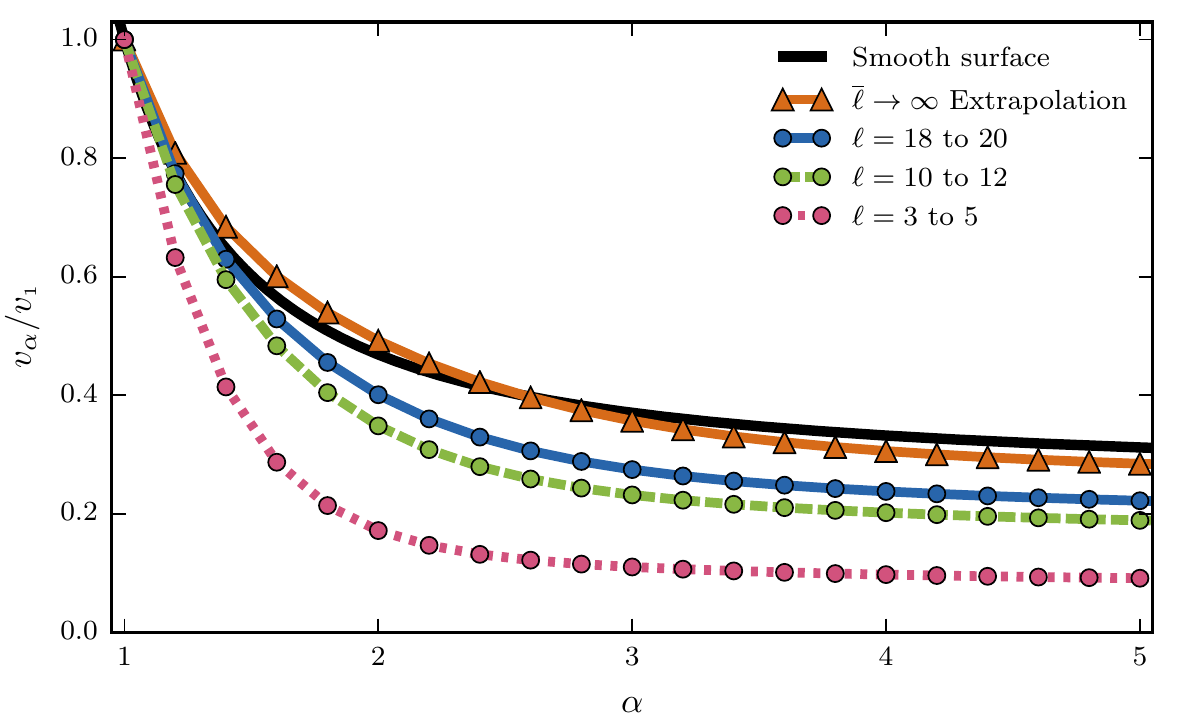}
	\end{center}
	\caption{The normalized logarithmic corner coefficient $v_{\alpha}/v_1$ as a function of the \Renyi index $\alpha$. 
	These results are based on the fitting procedure illustrated in Fig.~\ref{fig:VAlpha_fits}.
	For smooth surfaces, this ratio of logarithmic coefficients is known to behave as in Eq.~\ref{game1}.
	}
	\label{fig:cornerCoeff}
\end{figure*}

We first investigate what happens if we perform fits of $\mathcal{P}_\alpha$ to the two-parameter function $v_{\alpha} \log \ell + d_\alpha$ (ignoring, for the moment, additional subleading terms).
In Fig.~\ref{fig:VAlpha_fits}, we illustrate such fits for $\alpha = 1$ and $\alpha=5$. 
We perform fits over various ranges of the cluster order $\ell$ and find that for $\alpha=1$, the extracted value of $v_1$ is quite stable when this range of $\ell$ values is varied --- indicating that the unknown subleading terms in Eq.~\eqref{eq:P_alpha_scaling} are already negligible for the cluster sizes used in our calculations.
However, for $\alpha >1$, the value of $v_{\alpha}$ increases significantly as higher orders $\ell$ are included in the fit and it is important to consider the effects of subleading terms.
Indeed, at least one source of additional finite-size scaling correction for $\alpha >1$ is known to arise from the conical singularity of the multi-sheeted
Riemann surface,\cite{Sahoo2016}
although the functional form of this correction is only known for $d=1+1$.\cite{Cardy_2010}
In Fig.~\ref{fig:cornerCoeff}, we show the results for $v_{\alpha}/v_1$ versus $\alpha$ as extracted from these various fits.

In order to approximate $v_{\alpha}$ in the thermodynamic limit $\ell \to \infty$, we study the behavior of $v_{\alpha}(\overline{\ell})$ versus $\overline{\ell}$, as illustrated in the insets of Fig.~\ref{fig:VAlpha_fits}.
Here $\overline{\ell}$ is a length scale that characterizes the orders $\ell$ used to extract $v_{\alpha}$ from the two-parameter fits described above.
We choose to define $\overline{\ell}$ as the average order --- such that, for instance, $\overline{\ell}=19$ for the case where orders $\ell = 18$ to 20 are used in the initial fit of $\mathcal{P}_\alpha$ to $v_{\alpha} \log \ell + d_\alpha$.
We could, however, use other definitions of $\overline{\ell}$ such as the minimum or maximum cluster order.
We then extract the behavior of $v_{\alpha}$ for $\overline{\ell} \to \infty$ by fitting $v_{\alpha}(\overline{\ell})$ to the three-parameter function $v_{\alpha}^\infty + p_\alpha/(\overline{\ell} + q_\alpha)$.
Here $v_{\alpha}^\infty$, $p_\alpha$ and $q_\alpha$ are (fitted) constants, where $v_{\alpha}^\infty$ corresponds to $v_{\alpha}$ in the thermodynamic limit and $q_\alpha$ reflects the ambiguity in the definition of $\overline{\ell}$ as described above.
This $\overline{\ell} \to \infty$ extrapolation procedure is used for all $\alpha >1$.
For $\alpha=1$, $v_{\alpha}(\overline{\ell})$ is well-converged as a function of $\overline{\ell}$ and we estimate $v_{\alpha}$ simply from the initial two-parameter fit using the highest orders available.

Fig.~\ref{fig:cornerCoeff} shows that as higher orders are used in the fits, the extracted normalized corner coefficient $v_{\alpha}/v_1$ as a function of $\alpha$ approaches the functional behavior predicted in Eq.~\reef{game1}. 
Extrapolating to the infinite-size limit as described above appears to provide good agreement with this functional form, although we are not able to strictly quantify the agreement due to unknown finite-size errors within the NLCE procedure.

Hence, at least qualitatively, the $\alpha$-dependence agrees with that in Eq.~\req{game1}. Further, the latter dependence is identical to the behaviour of the universal coefficient appearing for a sphere, \ie $u_{\alpha}(S^2)/u_1(S^2)
=(\al+1)(\al^2+1)/(4\al^3)$. Hence this aspect of our results matches the observation in Ref.~\onlinecite{Devakul2014}, but we would also like to know how the overall coefficient in $v_{\alpha}$ relates to that of the spherical boundary, $u_\alpha(S^2)/8=-f_a(\al)/2$.  Our result for $v_1$ reads
\beq
v_1=+0.00286\, ,
\eeq
which differs both in magnitude and sign from the coefficient for an eighth of a sphere, \ie $u_1(S^2)/8=-1/720\simeq -0.00139$.
More generally, if we fit our results for $v_{\alpha}$ versus $\alpha$ to the form $\xi\, u_\alpha(S^2)/8$ for constant $\xi$, 
we find that 
\beq \label{sss}
v_{\alpha} \simeq \xi \, \left[-\frac12\,f_a(\al)\right]\, ,\quad \text{with } \xi=-2.06\, .
\eeq
Hence, our trihedral corner result presents a very similar $\alpha$-dependence, but differs significantly in magnitude, and also in sign, from the result corresponding to an eighth of the sphere.

\section{Discussion} \label{discuss}

In this paper, we have performed numerical linked-cluster calculations to evaluate the universal coefficient $v_\al$ produced by a cubic trihedral corner in  the \Renyi entropy of a massless free scalar field in $3+1$ dimensions. 
Our results suggest that the dependence of this coefficient on the \ren index $\al$ is well approximated by $v_{\alpha}/v_1 = (\alpha+1)(\alpha^2+1)/(4\alpha^3)$, which is the functional form expected for the universal coefficient appearing for a smooth spherical entangling surface.  
However, despite this similarity, our numerical results demonstrate that the magnitude and the sign of the universal coefficient for an eighth of a sphere does not match $v_{\alpha}$, as had been suggested in Ref.~\onlinecite{Devakul2014}.

The $\alpha$-dependence of the universal coefficient for general CFTs in the case of smooth surfaces is controlled by two independent functions, $f_a(\alpha)$ and $f_b(\alpha)$. In the special case of a massless free scalar, these functions are proportional to each other, as shown in Eq.~\req{name}.  
We have attempted to express the trihedral vertex coefficient $v_{\alpha}$ in terms of these two functions by using the general result valid for smooth surfaces in Eq.~\req{gamer} with a smoothed model of a cube in Section \ref{Scube}.  
However, this calculation gave a result that was problematic in two respects. 
First, it contained an unphysical $L/\delta\,\log( L/\delta)$ contribution, and second, the coefficient of the logarithmic term was regulator dependent. 
As noted, we expect that these problems arise since we have not properly accounted for the $\mathcal{O}((L/\delta)^0)$ terms in Eq.~\reef{game1hh}. 
In the singular limit $\veps\to\gamma_1\delta/L$, some of these overlooked contributions build up to restore the desired universality of the logarithmic coefficient and to eliminate the unphysical $L/\delta\,\log(L/\delta)$ term.

The phenomenon where lower order (\ie less divergent) contributions in the \Renyi entropy can build up to produce universal terms with a stronger divergence in a limit where the entangling surface becomes singular has been explicitly seen in Ref.~\onlinecite{Bueno4}. In particular, this effect was described for the appearance of a sharp corner in $2+1$ dimensions, and of a conical singularity in $3+1$ dimensions.
In the first case, small deformations of a circular entangling surface produce a universal contribution to the entanglement entropy that scales as $(L/\delta)^0$.\cite{Mezei14} 
These universal contributions can be evaluated for each Fourier mode on the circle, and when the Fourier modes are combined to produce a sharp corner, the sum of these contributions gives rise to a $\log(\ell/\delta)$ term.\cite{Bueno4}$^,$\footnote{These calculations are easily extended beyond $\al=1$ to general \Renyi entropies using the techniques introduced in Ref.~\onlinecite{displace}.} 
In the second example, the logarithmic contribution in Eq.~\req{gamer} becomes a $\log^2(\ell/\delta)$ term when a conical singularity appears in an otherwise smooth spherical entangling surface. 

However, there is an important difference between these two situations and the case of the trihedral corner considered here. 
Both for the  corner in $2+1$ dimensions and for the cone in $3+1$ dimensions, the order of the divergence corresponding to the universal term is higher for the singular surface than for the initial smooth surface. Hence, in the two cases studied in Ref.~\onlinecite{Bueno4}, the singular deformation changes the order of the divergence corresponding to the universal contribution, and hence the lower order contributions which are building up are in fact the universal contributions for the smooth entangling surfaces. Interestingly, this is not what happens for the trihedral corner, whose universal contribution is logarithmic, just like for a smooth entangling surface. Therefore, while it is natural to expect that some dependence on $f_{a}(\alpha)$ and $f_b(\alpha)$ survives in $v_{\alpha}$, it is also plausible that new contributions hidden in the $\mathcal{O}((L/\delta)^0)$ terms contribute.

It would be interesting to evaluate $v_{\alpha}$ for other CFTs in which $f_a(\alpha)$ and $f_b(\alpha)$ are independent (\ie rather than being proportional to one another, as in the free scalar theory) to gain a better understanding of the universal character of $v_{\alpha}$. On the one hand, the above discussion suggests that it may be premature to think that $v_{\alpha}$ is fully controlled by $f_a(\alpha)$ and $f_b(\alpha)$ alone. On the other hand, the $\al$ dependence of our numerical results is consistent with the trihedral coefficient being a simple linear combination
\beq
v_\al = \beta_a\,f_a(\al)+\beta_b\,f_b(\al)\,,
\label{linear}
\eeq
where $\beta_{a,b}$ are unspecified constants. Hence let us examine the latter possibility further.  First, we substitute the relation that $f_b(\al)=3\,f_a(\al)$, which holds for a massless free scalar, into Eq.~\reef{linear}. Then combining the resulting expression with the fit in Eq.~\reef{sss} yields
\beq
\beta_a+3\,\beta_b\simeq 1.03\,.
\eeq
Further, we recall that $f_a(\al)$ multiplies a topological term in Eq.~\reef{gamer}. If we assume that the topological nature of the $f_a(\al)$ contribution survives for the trihedral coefficient, we would find
\beq
\beta_a=-\frac12\,,\qquad\beta_b\simeq0.51\,.
\label{guess}
\eeq
The numerical value of the second coefficient is remarkably close to being $1/2$ and hence these somewhat speculative steps are pointing to a rather simple result:
\beq
v_\al \overset{?}{=} \frac12\big[f_b(\al)-f_a(\al)\big]\,.
\label{guess2}
\eeq
Using the functions in Eq.~\reef{name} for the massless free scalar, the above expression predicts
\beq
{\rm free\ scalar:}\quad v_\al\overset{?}{=} \frac{(\al+1)(\al^2+1)}{1440\al^3}\,.
\eeq
Now the simplest CFT in which $f_a(\alpha)$ and $f_b(\alpha)$ are not proportional to one another is a free Weyl fermion. In this theory, the two functions are given by\cite{Fursaev:2012mp,Lee:2014xwa}$^,$\footnote{As in the case of the scalar, the result for $f^\mt{Weyl}_b(\al)$ in Eq.~\reef{name2} relies on the equality $f_b(\al)=f_c(\al)$. Ref.~\onlinecite{Lee:2014xwa} provided strong numerical evidence that this equality holds for the free fermion, but no analytic proof has yet been constructed  in this case.}
\beqa
f^\mt{Weyl}_a(\al)&=&\frac{(\al+1)(37\al^2+7)}{5760\al^3}\, ,
\label{name2}\\
f^\mt{Weyl}_b(\al)&=&\frac{(\al+1)(17\al^2+7)}{1920\al^3}\, .
\nonumber
\eeqa
Combining these expressions with Eq.~\reef{guess2} then yields the surprisingly simple prediction:
\beq
{\rm Weyl\ fermion:}\quad v_\al \overset{?}{=} \frac{7}{4}\,\frac{(\al+1)(\al^2+1)}{1440\al^3}\,.
\label{name3}
\eeq
While this prediction relies on a number of unproven steps, it yields a very satisfying result. Namely, that the $\al$-dependence of $v_\al/v_1$ is {\it identical} for the free scalar and for the free fermion. We are currently extending our calculations to evaluate the trihedral corner coefficient $v_\al$ for the free fermion. 
We expect that with the numerical accuracy achieved here, it will be straightforward to distinguish the scaling in Eq.~\reef{name3} from, \eg that of $f^\mt{Weyl}_a(\al)$ or $f^\mt{Weyl}_b(\al)$ alone.  
Matching Eq.~\reef{name3} would provide strong evidence  that $v_{\alpha}$ is fully determined by $f_a(\alpha)$ and $f_b(\alpha)$, and by Eq.~\reef{guess2} in particular. 
On the other hand, disproving Eq.~\reef{name3} would suggest the trihedral corner provides new universal information beyond these two functions.

\begin{acknowledgments}   
We are thankful to Grigory Bednik, Horacio Casini, Johannes Helmes, Veronika Hubeny,
Bohdan Kulchytskyy, Max Metlitski, Sharmistha Sahoo, and especially William Witczak-Krempa for stimulating discussions.  
Research at Perimeter Institute is supported by the Government of Canada through Industry Canada and by the Province of Ontario through the Ministry of Research \& Innovation. LEHS gratefully acknowledges funding from the Ontario Graduate Scholarship. The work of PB was supported by a postdoctoral fellowship from the Fund for Scientific Research -- Flanders (FWO). 
PB also acknowledges support from the Delta ITP Visitors Programme. PB is grateful to the organizers of the ``It from Qubit Summer School'' held at the Perimeter Institute for Theoretical Physics and to the organizers of the ``Quantum Matter, Spacetime and Information'' conference held at the Yukawa Institute for Theoretical Physics (YITP) at Kyoto University.  
The work of RRPS is supported by the US National Science Foundation
grant number DMR-1306048.
RGM is supported in part by funding from the Natural Sciences and Engineering Research Council of Canada (NSERC) and a Canada Research Chair.
RCM is supported in part by funding from NSERC, from the Canadian Institute for Advanced Research and from the Simons Foundation through the ``It from Qubit'' collaboration.  
\end{acknowledgments} 

\appendix
\section{Edge entanglement } 
\label{app_edge}
In this appendix,  we present numerical calculations of the contribution to the \Renyi entropy produced by the presence of a $90$-degree wedge in the entangling surface. 
As discussed in Section~\ref{sec:renn}, we expect this non-universal contribution to be of the form $S_{\alpha,\text{edge}}=w_{\alpha}\, f(L/\delta)$ with $f(L\delta)=L/\delta$ where $L$ is the length of the edge and $\delta$ is the UV cutoff --- see Fig. \ref{fig1}. 
We examine this claim here and, in particular, we compare results corresponding to fitting our numerical results to this functional dependence against fits to $f(L/\delta)=(L/\delta)\log (L/\delta)$, which is the dependence appearing in our smoothed-cube calculation in Section~\ref{sec:renn}.

In these calculations, we forego the NLCE and instead study directly the behavior of the full $S_\alpha(A)$ for various regions $A$. In particular, we use the methods outlined in Section~\ref{sec:PeschelTrick} to calculate $S_\alpha(A)$ for cases where the full system is an $L \times L \times L$ cubic lattice (\textit{i.e.} $L_x=L_y=L_z\equiv L$) and subregion $A$ comprises an $L/2 \times L/2 \times L$ quadrant of the system (with $L$ even). 
Now we expect the entropies to scale as in Eq.~\eqref{eq:S_octant}, but without the logarithmic term (since the entangling surface $\partial A$ does not contain any corners). That is, 
\begin{equation}
S_{\alpha,\text{quad}}(L) = B_\alpha  2L^2  + 4\,S_{\alpha,\text{edge}} + \mathcal{O}(1),
\label{eq:S_quadrant}
\end{equation}
where $2L^2$ is the area of $\partial A$, and $S_{\alpha,\text{edge}}$ appears with a factor of 4 since the boundary $\partial A$ has 4 distinct edges.  
We perform our calculations on systems with either PBC or APBC along each direction. 
Further, as in Section~\ref{sec:Methods}, we measure $L$ in units of the lattice spacing, which we set to one for simplicity, \ie $\delta=1$.
Recall that, in order to consider a critical system, we examine a scalar field with zero mass. 

Due to translational invariance along the $z$ direction, we can access $S_{\alpha,\text{quad.}}$ on much larger lattices by utilizing the approach discussed, \eg, in Refs.~\onlinecite{Huerta:2011qi} and \onlinecite{Chen2014}.
Specifically, we perform a Fourier transform along $z$ in order to map our $(3+1)$-dimensional Hamiltonian to a set of $L$ separate $(2+1)$-dimensional models with effective masses that depend on the momenta $k_z$. 
We then use the methods of Section~\ref{sec:PeschelTrick} to calculate the \Renyi entropies for each of these $(2+1)$-dimensional ``slices" and then sum these contributions to get the overall $S_{\alpha,\text{quad.}}$ corresponding to the original $(3+1)$-dimensional model.

To compare the functional forms $f(L) = L$ and $f(L) =L\, \log L$ for $S_{\alpha,\text{edge}}$, we perform least-squares fits of our numerical data for $S_{\alpha,\text{quad}}(L)$ to the three-parameter function $2B_\alpha L^2 + w_\alpha f(L) + d_\alpha$.
We quantify the goodness of the fit by calculating the error
\begin{equation}
\Delta_\alpha = \sum_{i=1}^{n_{\scriptscriptstyle L}} \left( S_{\alpha,\text{quad}}(L_i) - \left[ 2B_\alpha L_i^2 + w_\alpha f(L_i) + d_\alpha \right] \right)^2,
\end{equation}
where $n_{\scriptscriptstyle L}$ is the number of values of $L$ used in the fit.

\begin{figure}[t]
	\begin{center}
		\includegraphics{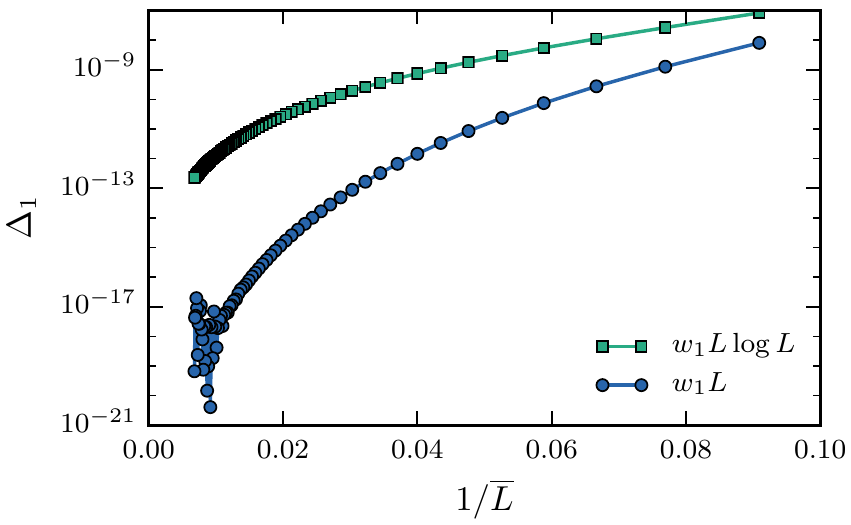}
	\end{center}
	\caption{The fitting errors $\Delta_1$ that result from fitting free boson data for $S_{1,\text{quad.}}(L)$ to the form $B_1 L^2 + w_1 f(L) + d_1$
		for $f(L) = L \log L$ and $f(L) = L$.
		We use $n_{\scriptscriptstyle L} = 4$ consecutive even values of $L$ for each fit, and define $\overline{L}$ to be the average of these 4 values.
		Note that the noise at high $\overline{L}$ is due to numerical issues of floating-point precision.
	}
	\label{fig:edgeFitErrors}
\end{figure}

Recall that in Section~\ref{sec:Results}, we found that the unknown subleading corrections to the entropies seem to be least significant for the von Neumann entropy, \ie $\alpha=1$. 
Hence for the present comparison, we focus on our data for $S_{1,\text{quad}}$. 
We perform the fits over several ranges of the lattice length $L$ with $n_{\scriptscriptstyle L} = 4$.
Fig.~\ref{fig:edgeFitErrors} illustrates results for the error $\Delta_1$ as a function of the average length $\overline{L}$ used in the fit. 
This plot utilizes PBC along the $x$ and $y$ lattice directions and APBC along $z$, and we find similar results for other combinations of PBC and APBC.
We conclude that, indeed, the linear function $f(L) = L$ provides a superior characterization of the edge contribution to the entropy since for large $L$, the corresponding errors $\Delta_1$ are several orders of magnitude lower than those corresponding to $f(L) = L\, \log L$. 

We also carry out this comparison for higher \Renyi entropies up to $\al=5$. 
We find that $\Delta_\alpha$ is significantly lower for the fits with $f(L)=L$ for all of these other values of $\alpha$. 
However, in these fits for $\alpha \neq 1$, we do not account for the unknown corrections that we expect produce a subleading $L$ dependence in the coefficient $w_\alpha$  (although these corrections should vanish in the thermodynamic limit).

Unlike the logarithmic coefficient coming from the trihedral corner, $w_\al$ is non-universal such it depends upon the procedure used to regulate the theory. 
A simple example of this regulator dependence comes from replacing $\delta$ by $2\delta$ in $S_{\alpha,\text{edge}}$, which changes $w_\al$ by a factor of two. 
One might expect this cutoff dependence to cancel out in the ratio $w_\alpha/w_1$ such that this ratio is universal. 
However, implicitly, this reasoning requires a covariant regulator. In our lattice computations, the area of the entangling surface $\partial A$ is only determined with a resolution of the lattice spacing such that ${\cal A}=2\ell^2 + \cO(\ell\delta)$ when $\ell$ is the characteristic length scale of region $A$. 
As a result, with small errors in $\cal A$, the leading area-law term generally ``pollutes'' the edge contribution, generating order one errors in the coefficient $w_\alpha$.\footnote{This effect is illustrated by our smoothed cube calculations in Section \ref{Scube}. There, $w_\al$ emerges in the limit $\veps\to\gamma_1\delta/L$ entirely from an $\cO(L\delta)$ contribution to the area in Eq.~\reef{area}. } 
Hence our lattice calculations cannot produce a reliable (\ie universal) result for the ratio $w_\alpha/w_1$. 
In principle, this issue can be evaded by performing the calculations with a covariant or geometric regulator.
Alternatively, a more careful treatment using mutual \Renyi information\footnote{Ref.~\onlinecite{Headrick:2010zt} defined the mutual \Renyi information (MRI) for two non-intersecting regions $A$ and $B$ as $I_\alpha(A,B)=S_\alpha(A)+S_\alpha(B)-S_\alpha(A\cup B)$, which is a natural generalization of the usual mutual information using \Renyi entropies. A particularly interesting feature of the MRI is that it is a finite regulator-independent quantity because all of the boundary divergences cancel.} allows one to produce reliable results with a lattice regulator. 
See, for instance, the discussion in Ref.~\onlinecite{chm2}.


\bibliography{References}{}    
\end{document}